\pdfoutput=1
\documentclass[
reprint,
superscriptaddress,
showpacs,
bibnotes,
amsmath,amssymb,
prb,
]{revtex4-1}

\usepackage{graphicx}
\usepackage{dcolumn}
\usepackage{bm}

\usepackage{setspace}
\begin{document}
\newcommand{\BFA}{BaFe$_{2}$As$_{2}$}
\newcommand{\TCS}{ThCr$_{2}$Si$_{2}$}
\newcommand{\LFAOF}{LaFeAs(O$_{1-x}$F$_x$)}
\newcommand{\TT}{$^{\circ}~2\theta$}
\newcommand{\SSOFA}{Sr$_{2}$ScO$_{3}$FeAs}
\newcommand{\SCOFA}{Sr$_{2}$CrO$_{3}$FeAs}
\newcommand{\SVOFA}{Sr$_{2}$VO$_{3}$FeAs}
\newcommand{\SVOFVA}{Sr$_{2}$VO$_{3}$(Fe$_{0.93}$V$_{0.07}$)As}
\newcommand{\SGOCS}{Sr$_{2}$GaO$_{3}$CuS}
\newcommand{\BaFA}{BaFe$_{2}$As$_{2}$}
\newcommand{\CSG}{$P\frac{4}{n}mm$}
\newcommand{\WP}{$wt$\%}

\title{Suppression of superconductivity by V-doping\\ and possible magnetic order in Sr$_{2}$VO$_{3}$FeAs}

\author{Marcus Tegel}%
\author{Tanja Schmid}%
\author{Tobias St\"{u}rzer}%
\author{Masamitsu Egawa}%
\affiliation{%
Department Chemie, Ludwig-Maximilians-Universit\"{a}t M\"{u}nchen, Butenandtstra\ss e 5-13 (Haus D), 81377 M\"{u}nchen, Germany\\}%

\author{Yixi Su}%
\affiliation{J\"{u}lich Centre for Neutron Science, IFF, Forschungszentrum J\"{u}lich, Outstation at FRM II, Lichtenbergstra{\ss}e 1, 85747 Garching, Germany}%

\author{Anatoliy Senyshyn}%
\affiliation{Institute for Materials Science, Darmstadt University of Technology, 64287 Darmstadt, Germany}%

\author{Dirk Johrendt}%
\email{johrendt@lmu.de}
\affiliation{Department Chemie, Ludwig-Maximilians-Universit\"{a}t M\"{u}nchen, Butenandtstra\ss e 5-13 (Haus D), 81377 M\"{u}nchen, Germany\\}%

\date{\today}

\begin{abstract}
Superconductivity at 33~K in Sr$_{2}$VO$_{3}$FeAs is completely suppressed by small amounts of V-doping in Sr$_{2}$VO$_{3}$(Fe$_{0.93(\pm0.01)}$V$_{0.07(\pm0.01)}$)As. The crystal structures and exact stoichiometries are determined by combined neutron- and x-ray powder diffraction. Sr$_{2}$VO$_{3}$FeAs is shown to be very sensitive to Fe/V mixing, which interferes with or even suppresses superconductivity. This inhomogeneity may be intrinsic and explains scattered reports regarding $T_c$ and reduced superconducting phase fractions in Sr$_{2}$VO$_{3}$FeAs. Neutron diffraction data collected at 4~K indicates incommensurate magnetic ordering of the V-sublattice with a propagation vector \textbf{q}\,$\approx$\,(0,0,0.306). This suggests strongly correlated vanadium, which does not contribute significantly to the Fermi surface of Sr$_{2}$VO$_{3}$FeAs.

\end{abstract}

\pacs{
74.70.Xa, 
74.62.Dh, 
74.62.En, 
61.05.fm, 
61.05.C-  
}

\maketitle




The discovery of iron pnictide superconductors \cite{Hosono-2008,Rotter-2-2008,Wang-2008} has opened a new chapter in superconductor research. Enormous progress has already been made with respect to the physics of these materials, and it becomes increasingly accepted that the weak magnetism inside the iron layers plays a decisive role in superconductivity,\cite{Mazin-2009-spins} even though the specific relationship to the pairing mechanism is still unclear. This weak magnetism appears as a spin-density-wave (SDW) in all iron based materials with simple PbFCl- \cite{Cruz-2008,Parker-2009} and ThCr$_2$Si$_2$-related structures,\cite{Rotter-1-2008} and is intimately connected with nesting of cylinder-shaped Fermi surfaces by a wave vector \textbf{q} = $(\pi, \pi)$.\cite{Mazin-2008-spm} However, no SDW of that or similar kind has been observed in the more complex arsenides Sr$_2M$O$_3$FeAs ($M$~=~Sc, Cr, V) \cite{Ogino-Sr2ScO3FeAs-2009,Tegel-21311,Wen-PhysRevB-2009} where the isoelectronic (FeAs)$^{1-}$ layers are separated by larger perowskite-like (Sr$_2M$O$_3$)$^{1+}$ blocks. Among them, only the V-compound \SVOFA~is superconducting up to 37~K \cite{Wen-PhysRevB-2009} and it has been controversially argued whether in this case the V-atoms significantly contribute to the Fermi surface \cite{Pickett-2010} or not \cite{Mazin-2010}. If this is not the case, the topology of the Fermi surface turns out to have the same essential features as in the other FeAs superconductors, otherwise one has to assume another mechanism in the case of \SVOFA. The latter seems to be improbable with respect to the $T_c$, which is strikingly similar to other iron arsenide superconductors.

If we assume for the moment that the superconducting mechanism in \SVOFA~is the same as in the other iron arsenides, the absence of any SDW anomaly may suggest that the FeAs layer is intrinsically doped. Vanadium can easily adopt oxidation states between V$^{1+}$ and V$^{4+}$ and would thus be able to supply or to accept electrons from the FeAs layers. Indeed, a recent x-ray absorption study indicates the presence of V$^{3+}$ and V$^{4+}$ in a \SVOFA~sample.\cite{V-21311-selfdoping} But even in this case, it remains confusing that also Sr$_2$ScO$_3$FeAs, where the scandium valence is fixed to Sc$^{3+}$, shows neither a SDW nor any other magnetic effect.\cite{Ogino-Sr2ScO3FeAs-2009}

However, thorough investigations of \SVOFA~are hampered by the poor quality of the samples, which always contain significant amounts of the ternary vanadium oxides Sr$_2$(VO$_4$), and/or Sr$_3$V$_2$O$_{7-\delta}$.\cite{Wen-PhysRevB-2009,V-21311-selfdoping,V-21311-singlestep} However, quantitative phase fractions are not specified by the authors. This might be dangerous in the present case, because these impurities contain V$^{4+}$ and exhibit at least weak paramagnetism.

Another fact that is hardly noticed so far concerns the true stoichiometry and homogeneity of the Sr$_2M$O$_3$FeAs compounds. The ionic radii of Fe$^{2+}$, V$^{2+/3+}$, and Cr$^{3+}$ are similar, and in face of synthesis temperatures above 1000$^{\rm{o}}$C, their mixing is easily conceivable. Indeed, we have recently found by neutron diffraction, that the chromium compound \SCOFA~is not stoichiometric, but intrinsically Cr-doped in the iron layer.\cite{tegel-epl-2010} We suggest that this mixing probably poisons superconductivity. Such mixing of iron and vanadium can also occur in \SVOFA, and already small amounts of V in the Fe layer may seriously affect the electronic and magnetic properties.

In order to shed some light on this puzzling material, we have optimized the synthesis of \SVOFA~in order to minimize the amounts of impurity phases, which allowed us to perform combined x-ray and neutron powder scattering investigations. In this paper, we report on the exact stoichiometry, magnetism and superconductivity of differently prepared \SVOFA-samples. We show that the superconducting phase is almost ideally stoichiometric, whereas small V-doping, which can easily be achieved in the Fe-layer, suppresses superconductivity. We present susceptibilty data and discuss hints to magnetic ordering of the V-sublattice from neutron scattering data.



\SVOFA~was synthesized by heating mixtures of strontium, vanadium, iron~(III)~oxide and arsenic~oxide in a molar ratio of 20:11:5:5. Two separate batches of 1 gram were prepared in alumina crucibles sealed in silica ampoules under an atmosphere of purified argon. Each mixture was heated to 1323~K at a rate of 60~K/h, kept at this temperature for 60~h and cooled down to room temperature. The products were homogenized in an agate mortar, pressed into pellets and sintered at 1323~K for 60 h. The latter step was performed twice. The batches were then united, reground, pressed into pellets of 6~mm in diameter and sintered together at 1323~K for 68~h. The obtained black crystalline product \SVOFA~is stable in air. \SVOFVA~was synthesized by heating mixtures of strontium, vanadium, iron~(III)~oxide, arsenic~oxide and vanadium~(V)~oxide in a ratio of 100:54:20:25:3 in two separate batches accordingly.

\begin{figure}[h]
\center{
\includegraphics[width=60mm]{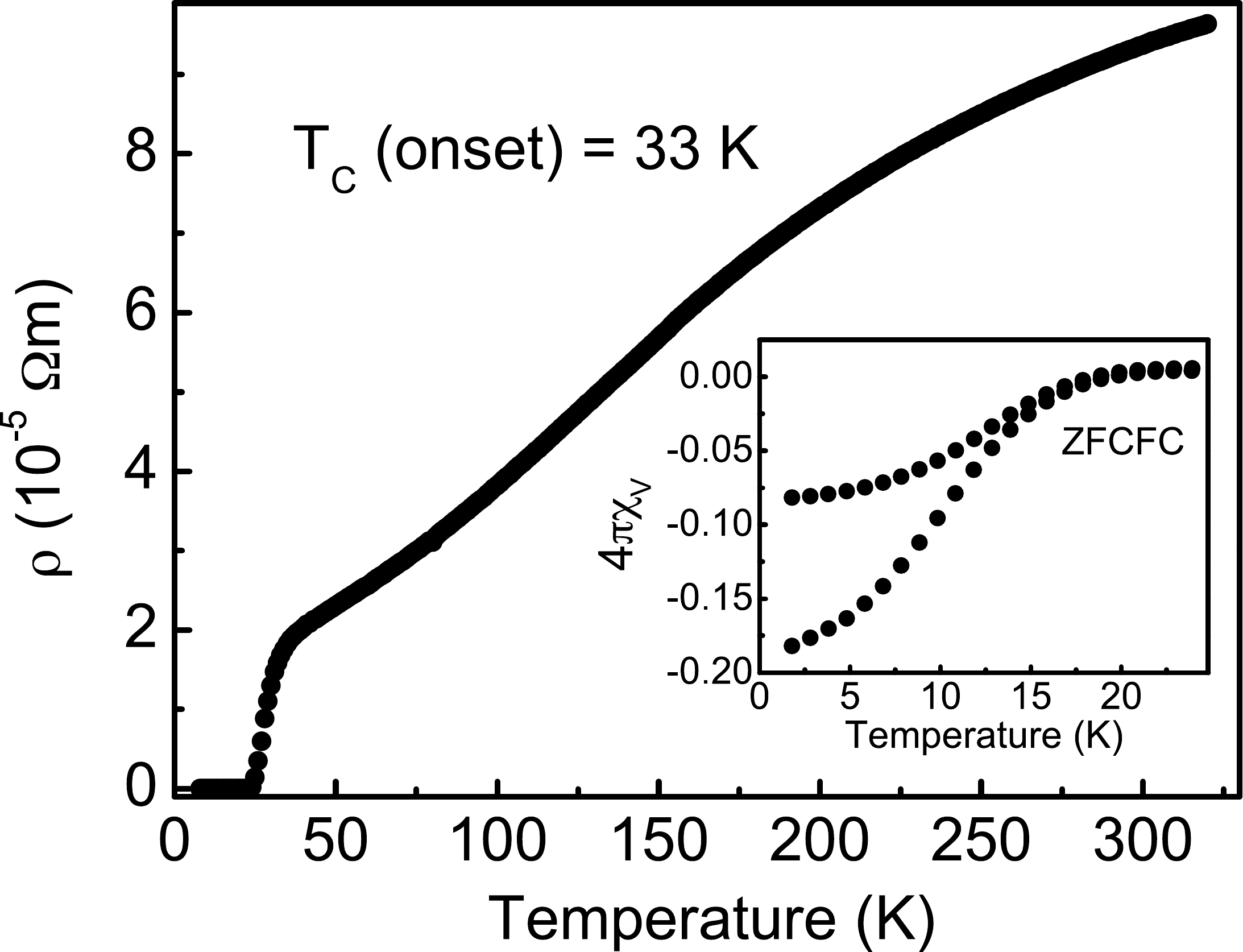}
\caption{Resistivity of \SVOFA. Inset: Zero-field-cooled / field-cooled measurements (20 Oe).}
\label{fig:Resistivity}
}
\end{figure}
\begin{figure}[h]
\center{
\includegraphics[width=70mm]{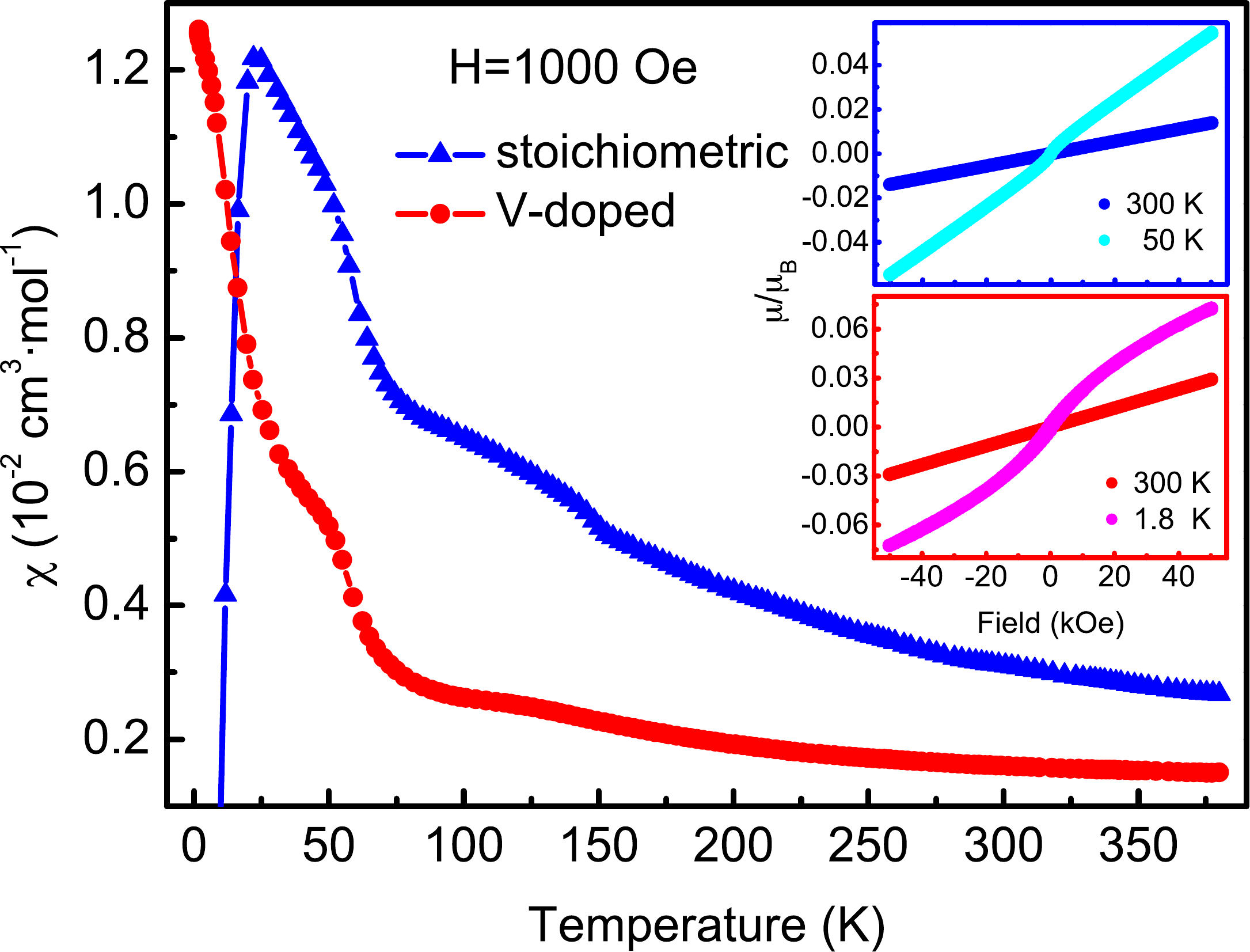}
\caption{(Color online) Molar susceptibility of \SVOFA ~(blue triangles) and \SVOFVA~(red circles) at 1~kOe. Insets: Hysteresis loops of \SVOFA~(blue) and \SVOFVA~(red).}
\label{fig:Magnetic}
}
\end{figure}

Resistivity measurements of the undoped sample show a rather broad superconducting transition at 33~K (Fig.~\ref{fig:Resistivity}). Superconductivity is verified by the zero-field-cooled field-cooled measurements using a Quantum Design MPMS-XL5 SQUID magnetometer, however, the estimated superconducting volume fraction is only 20\%. No superconductivity is detected in the V-doped sample. The susceptibilities of V-doped and undoped \SVOFA~measured at 1000~Oe and hysteresis loops at different temperatures are depicted in Fig.~\ref{fig:Magnetic}. Both samples exhibit Curie-Weiss-like paramagnetic behavior between 160 and 390~K. Anomalies appear at $\approx$~150~K, $\approx$~70~K and $\approx$~50~K in the stoichiometric sample, and similar also in the V-doped sample. Almost the same behavior has recently been reported in Ref.~\cite{V-21311-selfdoping}, where the authors suggest possible magnetic transitions of the Sr$_2$VO$_3$-layers and the absence of a structural transition, both in agreement with our neutron scattering results (see below).

X-Ray powder diffraction patterns at room temperature were recorded using a STOE STADI P diffractometer (Cu K$_{\alpha1}, \lambda$ = 0.154056~nm). Neutron powder diffraction patterns at 300~K and 4~K were recorded at the high resolution powder diffractometer SPODI at FRM II (Garching, Germany) with incident wavelengths of 0.155~nm and 0.146~nm, respectively. Rietveld refinements were performed with the TOPAS package \cite{topas} using the fundamental parameter approach as reflection profiles. Vanadium was used as a sample container for the neutron measurements and had to be included in the refinements. In order to describe small peak half width and shape anisotropy effects of the samples, the approach of Le Bail and Jouanneaux\cite{lbj} was implemented into the TOPAS program and the according parameters were allowed to refine freely. Preferred orientation of the crystallites were described using March Dollase or spherical harmonics functions. The Fe:V ratio at both the iron and the vanadium site was determined by refining the occupancy of the neutron and/or x-ray powder data, oxygen deficiency was ruled out by refining the occupancy of all oxygen sites.



\begin{figure}[h]
\center{
\includegraphics[width=70mm]{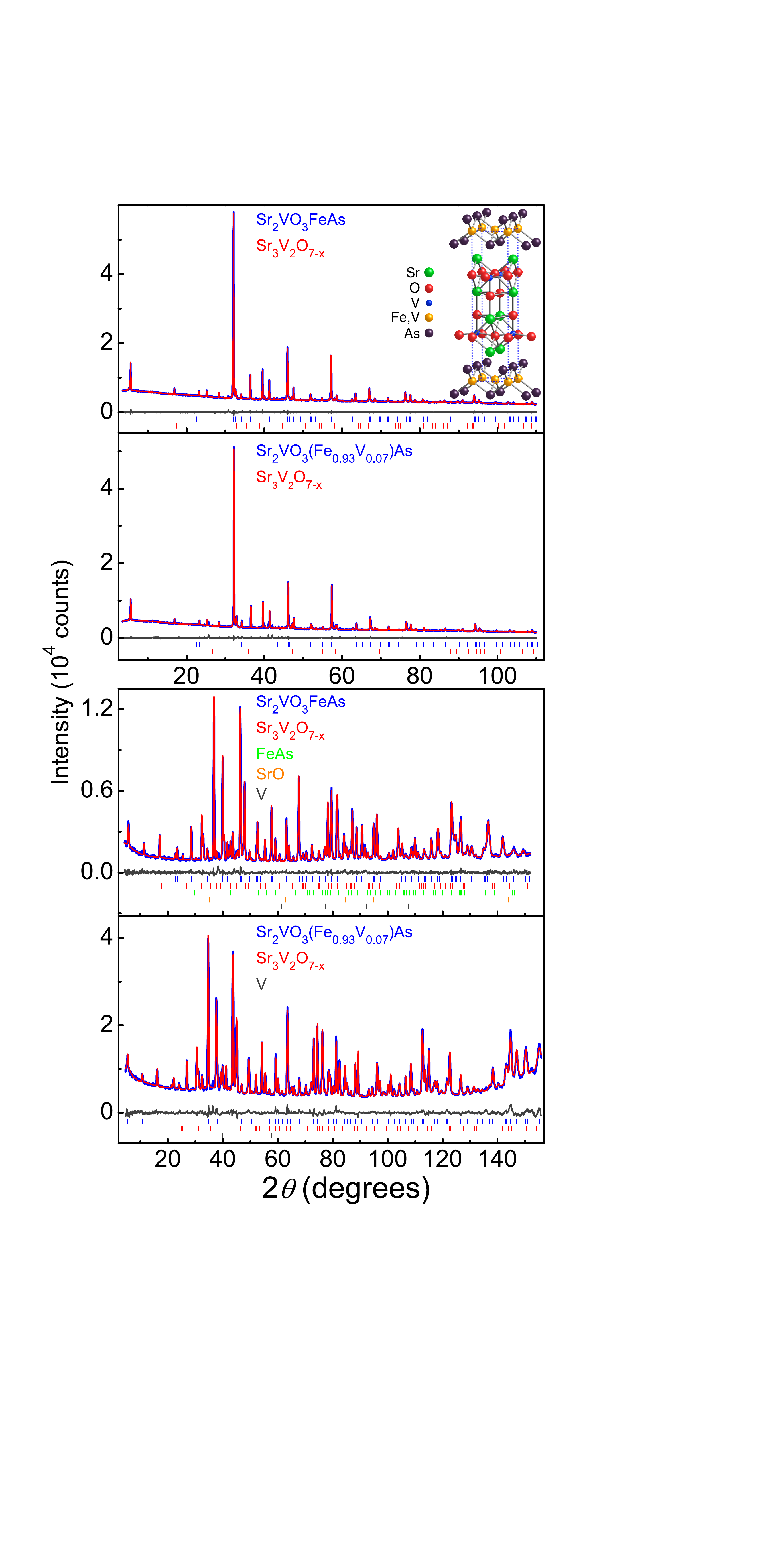}
\caption{(Color online) X-Ray (top) and neutron (bottom) powder patterns of \SVOFA~and \SVOFVA~with Rietveld refinements. Inset: Crystal structure of \SVOFA.}
\label{fig:Rietveld}
}
\end{figure}

The crystallographic data of \SVOFA~and \SVOFVA~are compiled in Table \ref{tab:Crystallographic}, the powder patterns and Rietveld fits are depicted in Fig. \ref{fig:Rietveld}. The amounts of impurity phases were determined by quantitative Rietveld analysis. The undoped sample consists of \SVOFA~(89.0 \WP), Sr$_{3}$V$_{2}$O$_{7-x}$ (7.8 \WP), FeAs (2.9 \WP) and traces of SrO (0.3 \WP), the doped sample consists of \SVOFVA~(87.3 \WP) and Sr$_{3}$V$_{2}$O$_{7-x}$  (12.7 \WP) as determined by neutron diffraction. However, the doped sample also shows small amounts of a further, unidentified impurity phase, which could not be included in the refinement. The lattice parameters change only slightly on V doping ($a$ is shortened by 1.2~pm, $c$ is unchanged), but the refinement of the Fe site (neutron data) displays a mixed occupancy of Fe and V in a ratio of $93\pm1$\%~Fe~:~$7\pm1$\%~V in the doped sample. Since the X-ray data gives full occupancy of the 2$a$-site (see~Table~\ref{tab:Crystallographic}), we can rule out vacancies at the Fe site and the Fe/V mixing is unambiguous. In contrast to this, the superconducting undoped sample is almost exactly stoichiometric regarding the V, Fe and O occupancies.

\begin{figure}[h]
\center{
\includegraphics[width=60mm]{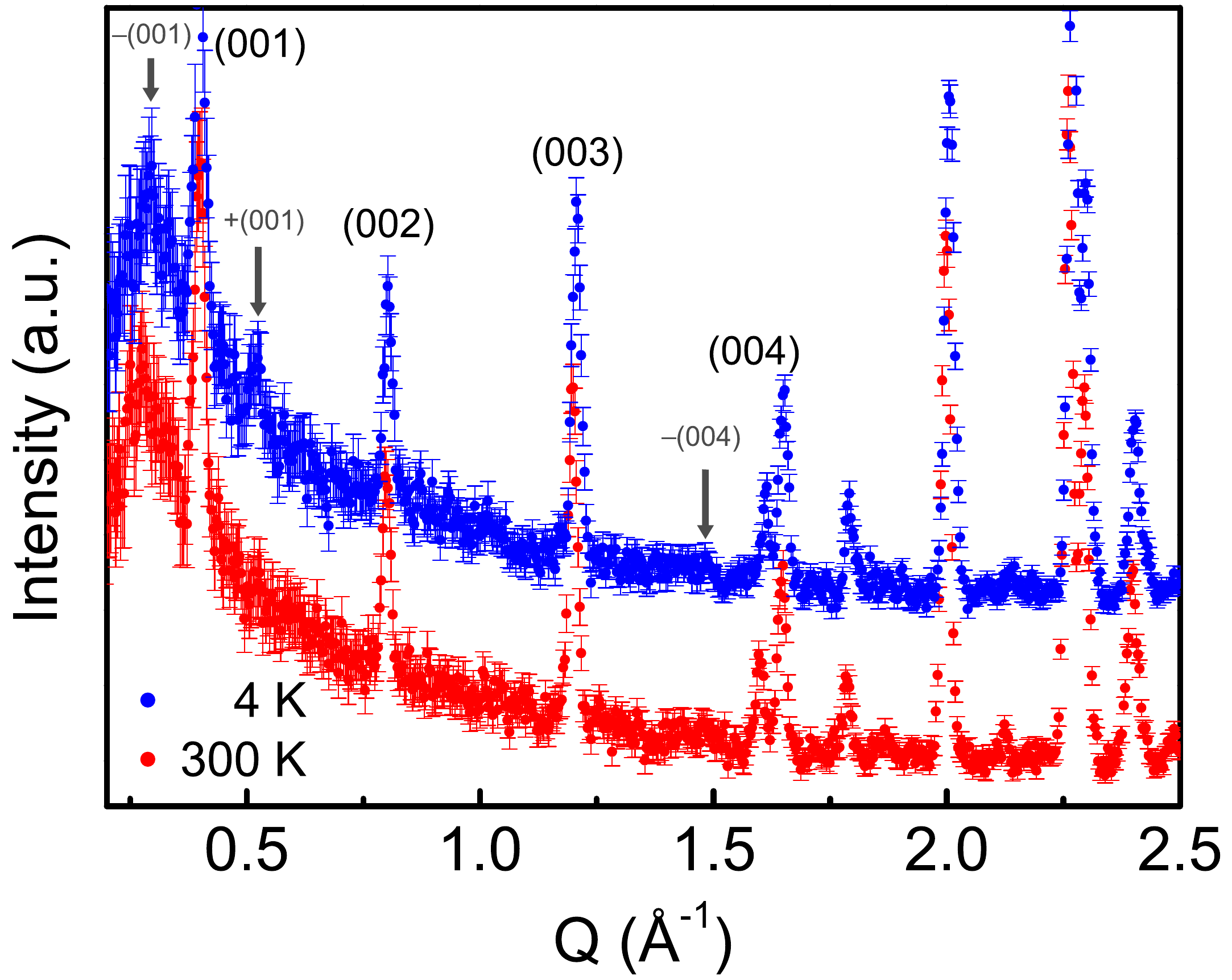}
\caption{(Color online) Neutron powder patterns of \SVOFA~ at 4~K (top) and 300~K (bottom). Magnetic reflections are marked arrows.}
\label{fig:NeutronMagnetic}
}
\end{figure}

The neutron powder pattern of undoped, superconducting \SVOFA~at 4~K shows weak additional peaks at the Q-values 0.525, 1.49 and possibly 0.28\,\AA$^{-1}$, marked by arrows in Fig.~\ref{fig:NeutronMagnetic}. These can be indexed as satellites of the (00l) reflections according to Q$_{(001)}+\Delta$Q, Q$_{(004)}-\Delta$Q and Q$_{(001)}-\Delta$Q with $\Delta$Q $\approx$ 0.123\,\AA$^{-1}$, respectively. This suggests incommensurate, possibly helical magnetic ordering along the \textit{c}-axis with a propagation vector \textbf{q} $\approx$ (0,0,0.307) r.l.u.. Since low-temperature $^{57}$Fe-M\"ossbauer-data show no signal splitting,\cite{V-21311-selfdoping} we expect ordering of the V-sublattice. This supports the idea of highly correlated vanadium in \SVOFA, where vanadium $d$-states are removed from the Fermi level by the magnetic exchange splitting. Recent angle resolved photoemission experiments \cite{Qian-2010} are in agreement with this model likewise. Even though our data strongly suggests the existence magnetic ordering, we consider them as preliminary. Further experiments in the low-Q region by polarization analysis are required for a precise determination of the magnetic structure and of the temperature dependence of the magnetic order parameter.

\begin{table*}
\caption{\label{tab:Crystallographic} Crystallographic data of \SVOFA~and \SVOFVA~at 300ÊK. Space group \CSG~(o1).}
\begin{ruledtabular}
\begin{tabular}{lllll}

								&\SVOFA					&					& \SVOFVA			&					 \\
\hline
diffractometer						& STOE (x-ray)			& SPODI (neutron)		& STOE (x-ray)			 & SPODI (neutron)		\\
wave length (nm) 					& 0.154					& 0.155				& 0.154				 & 0.146				 \\
\textit{a} (pm)\footnote{separate refinements without anisotropy paramters}& 394.58(1)				 & 393.5(1)			& 393.38(1)			&393.1(1)			\\
\textit{c} (pm)\footnotemark[1]			& 1573.1(1)				& 1570(1)				& 1573.1(1)			 & 1573(1)			\\
\textit{V} (nm$^{3}$)\footnotemark[1]	& 0.2449(1)				& 0.243(1)			& 0.2434(1) 			 & 0.243(1)			\\
\textit{Z} 							& 2						& 2					& 2					 & 2					 \\
data points 						& 10650					& 2955				& 10650				 & 3029				 \\
reflections\footnote{main phase}		& 130					& 196				& 130				 & 238				 \\
total variables 						& 101					& 95					& 94					 & 113				 \\
\textit {d} range 					& $0.941- 15.735$			& $0.797 - 15.700$		& $0.939-15.717$		 & $0.749-15.719$		\\
R$_P$, \textit{w}R$_P$ 				& 0.0147, 0.0196			& 0.0317, 0.0412		& 0.0187, 0.0264		 & 0.0321, 0.0405		\\
R$_{Bragg}$ 						& 0.0050					& 0.0101				& 0.0044				 & 0.0231				\\
\\

Atomic parameters: \\
\hline	
Sr1 [2$c$ ($0,\frac{1}{2},z$)] 			&$z$ = 0.8100(1)			& $z$ = 0.8096(1)													 &$z$ = 0.8096(2)			&$z$ = 0.8092(2)					\\
 								&$U_{iso} = 131(5)$			& $U_{iso} = 39(5)$													 &$U_{iso} = 171(6)$		&$U_{iso} = 49(7)$					\\
Sr2 [2$c$ ($0,\frac{1}{2},z$)]			&$z$ = 0.5859(1)			& $z$ = 0.5858(2)													 &$z$ = 0.5862(2)			&$z$ = 0.5852(2)					\\
 								&$U_{iso} = 220(7)$			& $U_{iso} = 96(6)$													 &$U_{iso} = 239(9)$		&$U_{iso} = 82(7)$					\\
V/Fe [2$c$ ($0,\frac{1}{2},z$)]			&$z$ = 0.3080(3)			& $z$ = 0.306(2)													 &$z$ = 0.3086(3)			& $z$ = 0.309(3)					\\
 								&$U_{iso} = 167(9)$			& $U_{iso} = 38$\footnote{restrained as minimum value}						&$U_{iso} = 179(11)$		&$U_{iso} = 25$\footnotemark[3]		 \\
								& 						& occ. 1.00(1):0.00(1)\footnote{restrained to a total occupancy $\leqslant 1$} 		& 						&occ. 1.00(1):0.00(1)\footnotemark[4]	\\
Fe/V [2$a$ ($0,0,0$)]				&$U_{iso} = 134(10)$		& $U_{iso} = 43(5)$													 &$U_{iso} = 131(13)$		&$U_{iso} = 40(6)$					\\
								&occ. 0.99(1)\footnotemark[4]	& occ. 0.99(1):0.01(1)\footnotemark[4]									 &occ. 1.00(1)\footnotemark[4]	 &occ. 0.93(1):0.07(1)\footnotemark[4]	\\
As [2$c$ ($0,\frac{1}{2},z$)]			&$z$ = 0.0891(2)			& $z$ = 0.0896(2)													 &$z$ = 0.0900(2)			&$z$ = 0.0901(2)					\\
 								&$U_{iso} = 113(7)$			& $U_{iso} = 34(6)$													 &$U_{iso} = 178(10)$		&$U_{iso} = 52(7)$					\\
O1 [4$f$ ($0,0,z$)] 					&$z$ = 0.2921(4)			& $z$ = 0.2933(1)													 &$z$ = 0.2958(5)			&$z$ = 0.2951(1)					\\
								&$U_{iso} = 93(19)$\footnote{$U_{iso}$ of O1 and O2 were refined together}	 & $U_{iso} = 56(5)$				&$U_{iso} = 305(31)$		&$U_{iso} = 65(7)$					 \\
								&occ. 1.00(1)\footnotemark[4]	& occ. 1.00(1)\footnotemark[4]											 &						&occ. 1.00(1)\footnotemark[4]			\\
O2 [2$c$ ($0,\frac{1}{2},z$)]			&$z$ = 0.4310(4)			& $z$ = 0.4297(1)													 &$z$ = 0.4313(7)			&$z$ = 0.4308(2)					\\
 								&$U_{iso} = 93(19)$\footnotemark[5]								& $U_{iso} = 94(8)$		  		&$U_{iso} = 419(54)$		&$U_{iso} = 131(10)$					 \\
								&occ. 1.00(1)\footnotemark[4]	& occ. 1.00(1)\footnotemark[4]											 &						&occ. 1.00(1)\footnotemark[4]			\\
\end{tabular}
\end{ruledtabular}
\end{table*}


In summary, we have shown that \SVOFA~is sensitive to Fe/V mixing in the FeAs layer, which is detrimental to superconductivity. Small V-doping of 7\%, unambiguously detected by neutron diffraction, suppresses superconductivity completely, while the superconducting phase is nearly stoichiometric. We suggest that even smaller Fe/V inhomogeneities are intrinsic in this material, and may be responsible for the scattered critical temperatures and superconducting phase fractions reported in the literature. Small but significant additional reflections emerge in the neutron powder pattern of superconducting \SVOFA~at 4~K.  A preliminary analyis indicates incommensurable, possibly helical magnetic ordering of the V-moments with a propagation vector \textbf{q}\,$\approx$\,(0,0,0.306). This is in agreement with strongly correlated vanadium, which does not significantly contribute to the Fermi surface. Thus, \SVOFA~fits to the other iron pnictide superconductors and represents no new paradigm, although the absence of iron magnetism and possible self doping effects remain open questions.

\begin{acknowledgments}

This work was financially supported by the German Research Foundation (DFG), Project No. JO257/6-1

\end{acknowledgments}


\begin{thebibliography}{19}
\expandafter\ifx\csname natexlab\endcsname\relax\def\natexlab#1{#1}\fi
\expandafter\ifx\csname bibnamefont\endcsname\relax
  \def\bibnamefont#1{#1}\fi
\expandafter\ifx\csname bibfnamefont\endcsname\relax
  \def\bibfnamefont#1{#1}\fi
\expandafter\ifx\csname citenamefont\endcsname\relax
  \def\citenamefont#1{#1}\fi
\expandafter\ifx\csname url\endcsname\relax
  \def\url#1{\texttt{#1}}\fi
\expandafter\ifx\csname urlprefix\endcsname\relax\def\urlprefix{URL }\fi
\providecommand{\bibinfo}[2]{#2}
\providecommand{\eprint}[2][]{\url{#2}}

\bibitem[{\citenamefont{Kamihara et~al.}(2008)\citenamefont{Kamihara, Watanabe,
  Hirano, and Hosono}}]{Hosono-2008}
\bibinfo{author}{\bibfnamefont{Y.}~\bibnamefont{Kamihara}},
  \bibinfo{author}{\bibfnamefont{T.}~\bibnamefont{Watanabe}},
  \bibinfo{author}{\bibfnamefont{M.}~\bibnamefont{Hirano}}, \bibnamefont{and}
  \bibinfo{author}{\bibfnamefont{H.}~\bibnamefont{Hosono}},
  \bibinfo{journal}{J. Am. Chem. Soc.} \textbf{\bibinfo{volume}{130}},
  \bibinfo{pages}{3296} (\bibinfo{year}{2008}).

\bibitem[{\citenamefont{Rotter et~al.}(2008{\natexlab{a}})\citenamefont{Rotter,
  Tegel, and Johrendt}}]{Rotter-2-2008}
\bibinfo{author}{\bibfnamefont{M.}~\bibnamefont{Rotter}},
  \bibinfo{author}{\bibfnamefont{M.}~\bibnamefont{Tegel}}, \bibnamefont{and}
  \bibinfo{author}{\bibfnamefont{D.}~\bibnamefont{Johrendt}},
  \bibinfo{journal}{Phys. Rev. Lett.} \textbf{\bibinfo{volume}{101}},
  \bibinfo{pages}{107006} (\bibinfo{year}{2008}{\natexlab{a}}).

\bibitem[{\citenamefont{Wang et~al.}(2008)\citenamefont{Wang, Liu, Lv, Gao,
  Yang, Yu, Li, and Jin}}]{Wang-2008}
\bibinfo{author}{\bibfnamefont{X.~C.} \bibnamefont{Wang}},
  \bibinfo{author}{\bibfnamefont{Q.~Q.} \bibnamefont{Liu}},
  \bibinfo{author}{\bibfnamefont{Y.~X.} \bibnamefont{Lv}},
  \bibinfo{author}{\bibfnamefont{W.~B.} \bibnamefont{Gao}},
  \bibinfo{author}{\bibfnamefont{L.~X.} \bibnamefont{Yang}},
  \bibinfo{author}{\bibfnamefont{R.~C.} \bibnamefont{Yu}},
  \bibinfo{author}{\bibfnamefont{F.~Y.} \bibnamefont{Li}}, \bibnamefont{and}
  \bibinfo{author}{\bibfnamefont{C.~Q.} \bibnamefont{Jin}},
  \bibinfo{journal}{Solid State Commun.} \textbf{\bibinfo{volume}{148}},
  \bibinfo{pages}{538} (\bibinfo{year}{2008}).

\bibitem[{\citenamefont{Mazin and Johannes}(2009)}]{Mazin-2009-spins}
\bibinfo{author}{\bibfnamefont{I.}~\bibnamefont{Mazin}} \bibnamefont{and}
  \bibinfo{author}{\bibfnamefont{M.~D.} \bibnamefont{Johannes}},
  \bibinfo{journal}{Nature Phys.} \textbf{\bibinfo{volume}{5}},
  \bibinfo{pages}{141} (\bibinfo{year}{2009}).

\bibitem[{\citenamefont{de~la Cruz et~al.}(2008)\citenamefont{de~la Cruz,
  Huang, and Lynn}}]{Cruz-2008}
\bibinfo{author}{\bibfnamefont{C.}~\bibnamefont{de~la Cruz}},
  \bibinfo{author}{\bibfnamefont{Q.}~\bibnamefont{Huang}}, \bibnamefont{and}
  \bibinfo{author}{\bibfnamefont{J.~W.} \bibnamefont{Lynn}},
  \bibinfo{journal}{Nature} \textbf{\bibinfo{volume}{453}},
  \bibinfo{pages}{899} (\bibinfo{year}{2008}).

\bibitem[{\citenamefont{Parker et~al.}(2009)\citenamefont{Parker, Pitcher,
  Baker, Franke, Lancaster, Blundell, and Clarke}}]{Parker-2009}
\bibinfo{author}{\bibfnamefont{D.~R.} \bibnamefont{Parker}},
  \bibinfo{author}{\bibfnamefont{M.~J.} \bibnamefont{Pitcher}},
  \bibinfo{author}{\bibfnamefont{P.~J.} \bibnamefont{Baker}},
  \bibinfo{author}{\bibfnamefont{I.}~\bibnamefont{Franke}},
  \bibinfo{author}{\bibfnamefont{T.}~\bibnamefont{Lancaster}},
  \bibinfo{author}{\bibfnamefont{S.~J.} \bibnamefont{Blundell}},
  \bibnamefont{and} \bibinfo{author}{\bibfnamefont{S.~J.}
  \bibnamefont{Clarke}}, \bibinfo{journal}{Chem. Commun.}
  \textbf{\bibinfo{volume}{16}}, \bibinfo{pages}{2189} (\bibinfo{year}{2009}).

\bibitem[{\citenamefont{Rotter et~al.}(2008{\natexlab{b}})\citenamefont{Rotter,
  Tegel, Schellenberg, Hermes, P\"ottgen, and Johrendt}}]{Rotter-1-2008}
\bibinfo{author}{\bibfnamefont{M.}~\bibnamefont{Rotter}},
  \bibinfo{author}{\bibfnamefont{M.}~\bibnamefont{Tegel}},
  \bibinfo{author}{\bibfnamefont{I.}~\bibnamefont{Schellenberg}},
  \bibinfo{author}{\bibfnamefont{W.}~\bibnamefont{Hermes}},
  \bibinfo{author}{\bibfnamefont{R.}~\bibnamefont{P\"ottgen}},
  \bibnamefont{and} \bibinfo{author}{\bibfnamefont{D.}~\bibnamefont{Johrendt}},
  \bibinfo{journal}{Phys. Rev. B} \textbf{\bibinfo{volume}{78}},
  \bibinfo{pages}{020503(R)} (\bibinfo{year}{2008}{\natexlab{b}}).

\bibitem[{\citenamefont{Mazin et~al.}(2008)\citenamefont{Mazin, Singh,
  Johannes, and Du}}]{Mazin-2008-spm}
\bibinfo{author}{\bibfnamefont{I.~I.} \bibnamefont{Mazin}},
  \bibinfo{author}{\bibfnamefont{D.~J.} \bibnamefont{Singh}},
  \bibinfo{author}{\bibfnamefont{M.~D.} \bibnamefont{Johannes}},
  \bibnamefont{and} \bibinfo{author}{\bibfnamefont{M.~H.} \bibnamefont{Du}},
  \bibinfo{journal}{Phys. Rev. Lett.} \textbf{\bibinfo{volume}{101}},
  \bibinfo{pages}{057003} (\bibinfo{year}{2008}).

\bibitem[{\citenamefont{Ogino et~al.}(2009)\citenamefont{Ogino, Katsura, Horii,
  Kishio, and Shimoyama}}]{Ogino-Sr2ScO3FeAs-2009}
\bibinfo{author}{\bibfnamefont{H.}~\bibnamefont{Ogino}},
  \bibinfo{author}{\bibfnamefont{Y.}~\bibnamefont{Katsura}},
  \bibinfo{author}{\bibfnamefont{S.}~\bibnamefont{Horii}},
  \bibinfo{author}{\bibfnamefont{K.}~\bibnamefont{Kishio}}, \bibnamefont{and}
  \bibinfo{author}{\bibfnamefont{J.}~\bibnamefont{Shimoyama}},
  \bibinfo{journal}{Supercond. Sci. Tech.} \textbf{\bibinfo{volume}{22}}
  (\bibinfo{year}{2009}).

\bibitem[{\citenamefont{Tegel et~al.}(2009)\citenamefont{Tegel, Hummel,
  Lackner, Schellenberg, P\"ottgen, and Johrendt}}]{Tegel-21311}
\bibinfo{author}{\bibfnamefont{M.}~\bibnamefont{Tegel}},
  \bibinfo{author}{\bibfnamefont{F.}~\bibnamefont{Hummel}},
  \bibinfo{author}{\bibfnamefont{S.}~\bibnamefont{Lackner}},
  \bibinfo{author}{\bibfnamefont{I.}~\bibnamefont{Schellenberg}},
  \bibinfo{author}{\bibfnamefont{R.}~\bibnamefont{P\"ottgen}},
  \bibnamefont{and} \bibinfo{author}{\bibfnamefont{D.}~\bibnamefont{Johrendt}},
  \bibinfo{journal}{Z. Anorg. Allg. Chem.} \textbf{\bibinfo{volume}{635}},
  \bibinfo{pages}{2242} (\bibinfo{year}{2009}).

\bibitem[{\citenamefont{Zhu et~al.}(2009)\citenamefont{Zhu, Han, Mu, Cheng,
  Shen, Zeng, and Wen}}]{Wen-PhysRevB-2009}
\bibinfo{author}{\bibfnamefont{X.~Y.} \bibnamefont{Zhu}},
  \bibinfo{author}{\bibfnamefont{F.}~\bibnamefont{Han}},
  \bibinfo{author}{\bibfnamefont{G.}~\bibnamefont{Mu}},
  \bibinfo{author}{\bibfnamefont{P.}~\bibnamefont{Cheng}},
  \bibinfo{author}{\bibfnamefont{B.}~\bibnamefont{Shen}},
  \bibinfo{author}{\bibfnamefont{B.}~\bibnamefont{Zeng}}, \bibnamefont{and}
  \bibinfo{author}{\bibfnamefont{H.~H.} \bibnamefont{Wen}},
  \bibinfo{journal}{Phys. Rev. B} \textbf{\bibinfo{volume}{79}},
  \bibinfo{pages}{220512} (\bibinfo{year}{2009}).

\bibitem[{\citenamefont{Lee and Pickett}(2010)}]{Pickett-2010}
\bibinfo{author}{\bibfnamefont{K.~W.} \bibnamefont{Lee}} \bibnamefont{and}
  \bibinfo{author}{\bibfnamefont{W.~E.} \bibnamefont{Pickett}},
  \bibinfo{journal}{EPL} \textbf{\bibinfo{volume}{89}}, \bibinfo{pages}{57008}
  (\bibinfo{year}{2010}).

\bibitem[{\citenamefont{Mazin}(2010)}]{Mazin-2010}
\bibinfo{author}{\bibfnamefont{I.}~\bibnamefont{Mazin}},
  \bibinfo{journal}{Phys. Rev. B} \textbf{\bibinfo{volume}{81}},
  \bibinfo{pages}{020507} (\bibinfo{year}{2010}).

\bibitem[{\citenamefont{Cao et~al.}(2010)\citenamefont{Cao, Ma, and
  Wang}}]{V-21311-selfdoping}
\bibinfo{author}{\bibfnamefont{G.}~\bibnamefont{Cao}},
  \bibinfo{author}{\bibfnamefont{Z.}~\bibnamefont{Ma}}, \bibnamefont{and}
  \bibinfo{author}{\bibfnamefont{X.}~\bibnamefont{Wang}},
  \bibinfo{journal}{arXiv:1007.3980}  (\bibinfo{year}{2010}).

\bibitem[{\citenamefont{Pal et~al.}(2009)\citenamefont{Pal, Vajpayee, Meena,
  Kishan, and Awana}}]{V-21311-singlestep}
\bibinfo{author}{\bibfnamefont{A.}~\bibnamefont{Pal}},
  \bibinfo{author}{\bibfnamefont{A.}~\bibnamefont{Vajpayee}},
  \bibinfo{author}{\bibfnamefont{R.~S.} \bibnamefont{Meena}},
  \bibinfo{author}{\bibfnamefont{H.}~\bibnamefont{Kishan}}, \bibnamefont{and}
  \bibinfo{author}{\bibfnamefont{V.~P.~S.} \bibnamefont{Awana}},
  \bibinfo{journal}{J. Supercond. Nov. Magn.} \textbf{\bibinfo{volume}{22}},
  \bibinfo{pages}{619} (\bibinfo{year}{2009}).

\bibitem[{\citenamefont{Tegel et~al.}(2010)\citenamefont{Tegel, Hummel, Su,
  Chatterji, Brunelli, and Johrendt}}]{tegel-epl-2010}
\bibinfo{author}{\bibfnamefont{M.}~\bibnamefont{Tegel}},
  \bibinfo{author}{\bibfnamefont{F.}~\bibnamefont{Hummel}},
  \bibinfo{author}{\bibfnamefont{Y.}~\bibnamefont{Su}},
  \bibinfo{author}{\bibfnamefont{T.}~\bibnamefont{Chatterji}},
  \bibinfo{author}{\bibfnamefont{M.}~\bibnamefont{Brunelli}}, \bibnamefont{and}
  \bibinfo{author}{\bibfnamefont{D.}~\bibnamefont{Johrendt}},
  \bibinfo{journal}{EPL} \textbf{\bibinfo{volume}{89}}, \bibinfo{pages}{37006}
  (\bibinfo{year}{2010}).

\bibitem[{\citenamefont{Coelho}(2007)}]{topas}
\bibinfo{author}{\bibfnamefont{A.}~\bibnamefont{Coelho}},
  \emph{\bibinfo{title}{TOPAS-Academic, version 4.1}}
  (\bibinfo{publisher}{Coelho Software, Brisbane}, \bibinfo{year}{2007}).

\bibitem[{\citenamefont{Le~Bail and Jouanneaux}(1997)}]{lbj}
\bibinfo{author}{\bibfnamefont{A.}~\bibnamefont{Le~Bail}} \bibnamefont{and}
  \bibinfo{author}{\bibfnamefont{A.}~\bibnamefont{Jouanneaux}},
  \bibinfo{journal}{J. Appl. Cryst.} \textbf{\bibinfo{volume}{30}},
  \bibinfo{pages}{265} (\bibinfo{year}{1997}).

\bibitem[{\citenamefont{Qian et~al.}(2010)\citenamefont{Qian, Xu, Shi,
  Nakayama, Richard, Kawahara, Sato, Takahashi, Neupane, Xu
  et~al.}}]{Qian-2010}
\bibinfo{author}{\bibfnamefont{T.}~\bibnamefont{Qian}},
  \bibinfo{author}{\bibfnamefont{N.}~\bibnamefont{Xu}},
  \bibinfo{author}{\bibfnamefont{Y.-B.} \bibnamefont{Shi}},
  \bibinfo{author}{\bibfnamefont{K.}~\bibnamefont{Nakayama}},
  \bibinfo{author}{\bibfnamefont{P.}~\bibnamefont{Richard}},
  \bibinfo{author}{\bibfnamefont{T.}~\bibnamefont{Kawahara}},
  \bibinfo{author}{\bibfnamefont{T.}~\bibnamefont{Sato}},
  \bibinfo{author}{\bibfnamefont{T.}~\bibnamefont{Takahashi}},
  \bibinfo{author}{\bibfnamefont{M.}~\bibnamefont{Neupane}},
  \bibinfo{author}{\bibfnamefont{Y.-M.} \bibnamefont{Xu}},
  \bibnamefont{et~al.}, \bibinfo{journal}{arxiv:1008.4905}
  (\bibinfo{year}{2010}).

\end{thebibliography}

\providecommand{\noopsort}[1]{}\providecommand{\singleletter}[1]{#1}%

\end{document}